\documentclass[prd, twocolumn, superscriptaddress,floatfix, nofootinbib, preprintnumbers]{revtex4-2}
\usepackage[utf8]{inputenc}
\usepackage[colorlinks=true,citecolor=blue]{hyperref}
\usepackage{amssymb}
\usepackage{amsmath}
\usepackage{graphicx}
\usepackage{footmisc}
\usepackage{url}
\usepackage{multirow}
\usepackage{makecell}
\usepackage{enumerate}
\usepackage{hhline}
\usepackage{comment}
\usepackage{nicefrac}
\usepackage{placeins}
\usepackage[separate-uncertainty=true]{siunitx}     % Unified number/unit handling
  \DeclareSIUnit\year{yr}
\usepackage[svgnames]{xcolor}
\usepackage[capitalize]{cleveref}
\usepackage{physics}
\usepackage{aas_macros}

\def\apjl{\rmfamily{ApJ~}}
\def\apjs{\rmfamily{ApJS~}}       % Astrophysical Journal, Supplement
\def\beq{\begin{equation}}
\def\eeq{\end{equation}}
\newcommand{\bea}{\begin{eqnarray}\begin{aligned}}
\newcommand{\eea}{\end{aligned}\end{eqnarray}}

\begin{document}

%%%%%%%%%%%%%%%%%%%%%%%%%%%%%%%%%%%%%%%%%%%%%%%%%%%%%%%%%%
\title{
  Fast Parameter Inference on Pulsar Timing Arrays with  Normalizing Flows
}
%%%%%%%%%%%%%%%%%%%%%%%%%%%%%%%%%%%%%%%%%%%%%%%%%%%%%%%%%%

%%%%%%%%%%%%%%%%%%%%%%%%%%%%%%%%%%%%%%%%%%%%%%%%%%%%%%%%%%
\author{David Shih}
\affiliation{New High Energy Theory Center, Rutgers University \\
  Piscataway, New Jersey 08854-8019, USA}

\author{Marat Freytsis}
\affiliation{LEPP, Department of Physics, Cornell University, Ithaca, NY 14853, USA}

\author{Stephen R.\ Taylor}
\affiliation{Department of Physics and Astronomy, Vanderbilt University,
2301 Vanderbilt Place, Nashville, TN 37235, USA}

\author{Jeff A.\ Dror}
\affiliation{Department of Physics, University of Florida, Gainesville, FL 32611, USA}
\affiliation{Department of Physics, University of California Santa Cruz, 1156 High St., Santa Cruz, CA
95064, USA}

\author{Nolan Smyth}
\affiliation{Department of Physics, University of California Santa Cruz, 1156 High St., Santa Cruz, CA
95064, USA}
%%%%%%%%%%%%%%%%%%%%%%%%%%%%%%%%%%%%%%%%%%%%%%%%%%%%%%%%%%

\begin{abstract}
  Pulsar timing arrays (PTAs) perform Bayesian posterior inference with expensive MCMC methods. Given a dataset of $\sim 10\text{--}100$ pulsars and ${\mathcal O}(10^3)$ timing residuals each, producing a posterior distribution for the stochastic gravitational wave background (SGWB) can take days to a week. The computational bottleneck arises because the likelihood evaluation required for MCMC is extremely costly when considering the dimensionality of the search space. Fortunately, generating simulated data is fast, so modern simulation-based inference techniques can be brought to bear on the problem. In this paper, we demonstrate how conditional normalizing flows trained on simulated data can be used for extremely fast and accurate estimation of the SGWB posteriors, reducing the sampling time from weeks to a matter of seconds.
\end{abstract}

\maketitle

\textbf{Introduction.} Pulsar timing array (PTA) experiments have recently announced evidence for an all-sky background of gravitational waves (GWs) in a frequency window of $\sim 1\text{--}100$ nHz \citep{NANOGrav:2023gor,EPTA:2023fyk,Reardon:2023gzh,Xu:2023wog}. These experiments leverage the exceptional timing regularity of millisecond pulsars to search for quadrupolar-like correlated arrival-time deviations of radio pulses, thereby signaling the presence of GWs. The origin of these GWs is still uncertain, but a known source in this frequency range is from a population of sub-parsec--separated supermassive black-hole binaries whose individual GW signals superpose incoherently to produce a stochastic background (see, e.g., \cite{NANOGrav:2023hfp,EPTA:2023xxk} and references therein). Additionally, there could be gravitational waves arising from new physics in the early Universe~\citep{NANOGrav:2023hvm}. The origin of this signal, and the underlying astrophysics and cosmology leading to it, will become better understood as existing pulsars are timed longer and more pulsars are added to arrays \citep{NANOGrav:2020spf}. Yet this will lead to existing PTA inference strategies becoming ever more taxed, lengthening analysis times and reducing the scope of studies that can be tackled with available computational resources. There are ongoing efforts to resolve this problem, including techniques that refit on intermediate analysis products and condense them into sufficient statistics \citep{2023arXiv230315442L}. However, neural posterior estimation and other deep learning techniques have not yet been used to confront PTA GW inference. 

One of the essential aims of PTA inference is to learn the posterior density $p(\theta|r)$
in parameter space (e.g., the amplitude and power-law slope of the SGWB signal) given the data $r$. The standard approach to deriving this posterior density is MCMC sampling of the true likelihood $p(r|\theta)$, which is modeled as a Gaussian distribution over the data. The covariance for individual pulsars is determined by red and white noise processes, while the inter-pulsar covariance is based on the Hellings--Down curve \citep{1983ApJ...265L..39H} of the SGWB.
To evaluate this likelihood, one must invert a large covariance matrix (with dimensions of the product of the number of pulsars and the number of times of arrival for each pulsar) for each parameter vector visited in an MCMC chain. Deriving a single posterior density for the NANOGrav 12.5 year dataset \citep{NANOGrav:2020gpb} requires $\gtrsim5$ days of computation to explore a Hellings--Downs-correlated model of the SGWB, while simultaneously sampling the intrinsic pulsar red-noise processes.\footnote{Likelihood evaluation times on this dataset for a Hellings--Downs-correlated model can be $\sim 0.1\text{--}1$ seconds. At least $10^6$ likelihood iterations are usually performed in an MCMC exploration of the model space, resulting in $\sim 5$ days of computational walltime.} 

This is an ideal situation for simulation based inference (SBI) (see, e.g., \cite{Cranmer_2020} for a review). The forward model (simulating time series from parameters) is extremely fast, since the time series have diagonal covariance in Fourier space, and the Fourier transform is fast to evaluate. In contrast, the likelihood evaluation, which necessitates inverting the covariance matrix, is time-consuming. This can be avoided by generating a large training dataset consisting of pairs of parameters and time series $(\theta,r)$; the way we sample over the parameter space forms the prior. The techniques of SBI can then be used to learn a fast-sampling posterior density $p(\theta|r)$ from the samples. SBI has already proven to be highly successful across a wide range of domains (see \cite{SBI_website_Cranmer} and \cite{SBI_website_Sharma} for curated and continuously updated bibliographies). Here we demonstrate for the first time the power of using SBI to analyze pulsar timing data. 

Specifically, we will show how a {\it normalizing flow} (NF) can learn the posterior density from samples in a matter of hours, and reduce the sampling time from days to a matter of seconds. Normalizing flows are a powerful method for density estimation and generative modeling (see~\citep{Kobyzev_2021,papamakarios2021normalizing} for reviews and original references). Using highly expressive neural networks they aim to learn an invertible transformation with tractable Jacobian between any data distribution to a latent space following a simple pre-specified distribution (such as Gaussian or uniform). By running this transformation in one direction one can estimate the probability density of any point in the data set; running it in other direction one can generate more samples that follow the same distribution as the data. The application of normalizing flows to learn posterior densities was first proposed in~\cite{papamakarios2018fast}. The purpose of this paper is to demonstrate the utility of normalizing flows in speeding up the inference of pulsar and SGWB parameters in pulsar timing arrays.

\textbf{Data.} Pulsars are modeled as having stationary pulse-arrival time series (i.e., random Gaussian processes fully characterized by their power spectra in frequency space), a good approximation when the gravitational wave frequencies are well above the inverse observation time of the pulsar~\cite{DeRocco:2023qae}. The relevant parameters are white noise; individual ``red noise'' for each pulsar $I$ with $A_r^{(I)}$, $\gamma_r^{(I)}$ describing the amplitude and exponent of a power-law model; and the SGWB which is common to all pulsars, with amplitude $A_\mathrm{GW}$ and $\gamma_\mathrm{GW}$, again for a power-law spectral model. If supermassive black hole mergers are the dominant source of gravitational waves in the nanohertz range, then it is expected that pulsar timing will observe $\gamma_\mathrm{GW} \approx 13/3$ \citep{2001astro.ph..8028P}. We will keep it as a free parameter in this study and seek to infer it from the data. 

Our goal is to build mock pulsar timing datasets that model the key sources of signal and noise. Although common software frameworks exist for pulsar timing analysis (most notably \texttt{libstempo}~\cite{2020ascl.soft02017V} and \texttt{PINT}~\citep{2021ApJ...911...45L}), we choose to generate our training data using our own code for greater simulation efficiency, control over the data, and understanding of the results. (For details of our methods, see \cref{app:homebrew}.) Using our own framework, we generate one million mock PTA residual time series for $N_p = 10$ pulsars with observation times drawn from the epoch-averaged NANOGrav 12.5~year dataset \citep{NANOGrav:2020gpb}. We use the 10 pulsars which were found to contribute the most evidence toward an (isotropic) SGWB signal in the NANOGrav analysis of their dataset~\cite{NANOGrav:2020bcs}. \Cref{tab:pulsarinfo} describes these 10 pulsars and their best-fit red noise parameters. The white noise is fixed to 100~ns for all pulsars; red noise is sampled independently for each pulsar, from a uniform distribution, $\log_{10} A^{(i)}_r \in [-19,-13]$, $\gamma^{(i)}_r \in [1,7]$; and the SGWB is sampled uniformly from $\log_{10} A_\mathrm{GW} \in [-18,-13]$, $\gamma_\mathrm{GW} \in [1,7]$. Red noise and SGWB contributions to the residuals are generated in frequency space and then Fourier transformed to the time domain; see \cref{app:homebrew} for details. We include a minimal pulsar timing model for each pulsar which accounts for a time-offset, the pulsar period, and the rate of change of the pulsar period. To remove the dependence on the Fourier transform base frequency and the pulsar timing model, we apply the $\mathsf{G}$-matrix projection, following~\cite{vanHaasteren:2012hj}. This corresponds to the marginalizing over the timing model parameters with a uniform prior on the deformations away from the true values. After the $\mathsf{G}$-matrix projection, there are a total of 2,940 projected residuals across the 10 pulsars. Of the $10^6$ generated time series, we reserve 90\% for training the normalizing flow (to be described in the next section) and 10\% for validation (model selection). 

We found that, in our generated data, the residuals spanned an enormous range---nearly 14 orders of magnitude---and could take either sign. This was a challenge to pre-process the residuals into a form that enabled the flow to learn effectively. We found rescaling the residuals, $r \to 10^7\times r$, followed by a clipping $\pm 1000$, worked well to make the inputs of the neural network ${\mathcal O}(1)$. This focuses on the part of the parameter space of greatest interest (the weakly-detectable SGWB regime) and might lose sensitivity to the part of parameter space of less interest (a huge SGWB signal, which is anyways incompatible with current observations). Finally, each time series $r \in \mathbb{R}^{2940}$ is paired with the set of parameters which characterize the stochastic noise, $\theta \in \mathbb{R}^{22}$; these parameters are:
\begin{multline}
  \theta = (\log_{10} A_\mathrm{GW}, \gamma_\mathrm{GW}, \\
            \log_{10} A_r^{(1)}, \gamma_r^{(1)}, \dotsc,
            \log_{10} A_r^{(10)}, \gamma_r^{(10)}) \,.
\end{multline}
Since the training data is generated with a uniform distribution, we preprocess $\theta$ with a simple shift and rescaling so $\theta \in [-1,1]^{22}$.

\begin{table}[t]

\begin{ruledtabular}
\begin{tabular}{cccc}
    Name & No.\ residuals & 
    Best-fit $\log_{10}A_{r}$ &
    Best-fit $\gamma_{r}$ 
    \\
    \hline
    J1909\texttt{-}3744 & 408 & -15.08 & 1.73\\
    J2317\texttt{+}1439 & 447 & -17.08 & 3.20\\
    J2043\texttt{+}1711 & 302 & -16.39 & 2.94\\
    J1600\texttt{-}3053 & 236 & -13.54 & 0.61\\
    J1918\texttt{-}0642 & 262 & -16.38 & 2.68\\
    J0613\texttt{-}0200 & 278 & -14.46 & 2.16\\
    J1944\texttt{+}0907 & 136 & -16.51 & 3.06\\
    J1744\texttt{-}1134 & 268 & -13.62 & 2.45\\
    J1910\texttt{+}1256 & 170 & -16.70 & 3.25\\
    J0030\texttt{+}0451 & 463 & -15.08 & 4.89\\
    \hline
\end{tabular}
\end{ruledtabular}
\caption{The 10 pulsars used in this analysis, their number of residuals, and best fit red noise parameters, $A_r$ and $\gamma_r$.
}
\label{tab:pulsarinfo}
\end{table}

\textbf{ML setup.} We fit a conditional normalizing flow to samples $(\theta,r)$ in order to estimate the posterior density $p(\theta|r)$. Our NF architecture is as follows. We use Masked Autoregressive Flows (MAFs) \cite{papamakarios2018masked} with Rational Quadratic Spline (RQS) transformations \cite{durkan2019neural,
10.1093/imanum/2.2.123}. The flow is comprised of 8 MADE blocks \cite{germain2015made}. Each MADE block has 2 hidden layers with 200 nodes each and ReLU activations. Each MADE block outputs the parameters of an RQS transformation with 8 bins and tail bound $B=1$. The base distribution is also taken to be uniform from $[-1,1]$. Our choices of tail bound and base distribution are motivated by the uniformly sampled training data; in early tests we found the flow performed better this way compared to using a Gaussian base distribution.

Rather than feeding all 2,940 residuals directly to the NF, we first pass them through an auxiliary embedding network. We found the following multi-stage architecture worked well:
\begin{equation}
  r' = F\big(E_1(r_1), E_2(r_2), \dotsc, E_{10}(r_{10})\big)\,.
\end{equation}
Here each $E_I$ takes the (preprocessed) residuals of pulsar $I$ and returns a per-pulsar embedding; then $F$ takes the concatenation of these embeddings and returns an overall embedding.  $E_I$ consists first of an LSTM \cite{sak2014long} with 100-dimensional hidden state and 2 hidden layers, which outputs the concatenation of the final hidden and cell state. These are then fed to a 2-layer multilayer perceptron (MLP) with 200, 100 nodes (ReLU activations) and a 50 node output layer (linear activation). Finally, $F$ is just a simple 2-layer MLP with 100 hidden nodes in each layer (ReLU activations) and a 50 node output layer (linear activation). We found that using an LSTM in the $E_I$ instead of just an MLP improved the performance of the network significantly. So did using a 2-stage per-pulsar structure instead of feeding all 2,940 residuals to a single LSTM or MLP.

To implement our normalizing flow and embedding network, we use the \texttt{nflows} package~\cite{conor_durkan_2020_4296287} and \texttt{PyTorch}~\cite{10.5555/3454287.3455008}. The entire setup (NF plus embedding network) is trained concurrently using the log-likelihood objective and the RAdam optimizer~\cite{liu2021variance} with default parameters. The networks are trained for up to 100 epochs, and the epoch with the best validation loss is chosen for the final demonstration. 
The training took approximately 24 hours on 3 Nvidia P100 GPUs. Sampling the flow to produce the posteriors takes a matter of seconds.

\begin{figure*}[t]
  \includegraphics[width=0.24\linewidth]{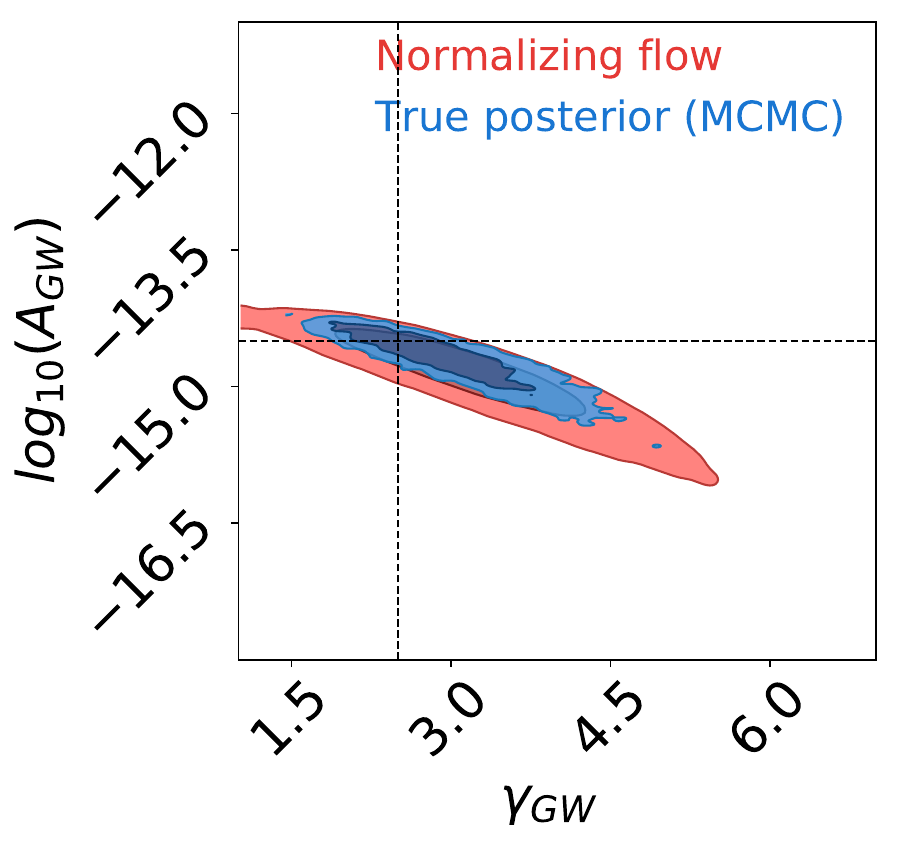}   \includegraphics[width=0.24\linewidth]{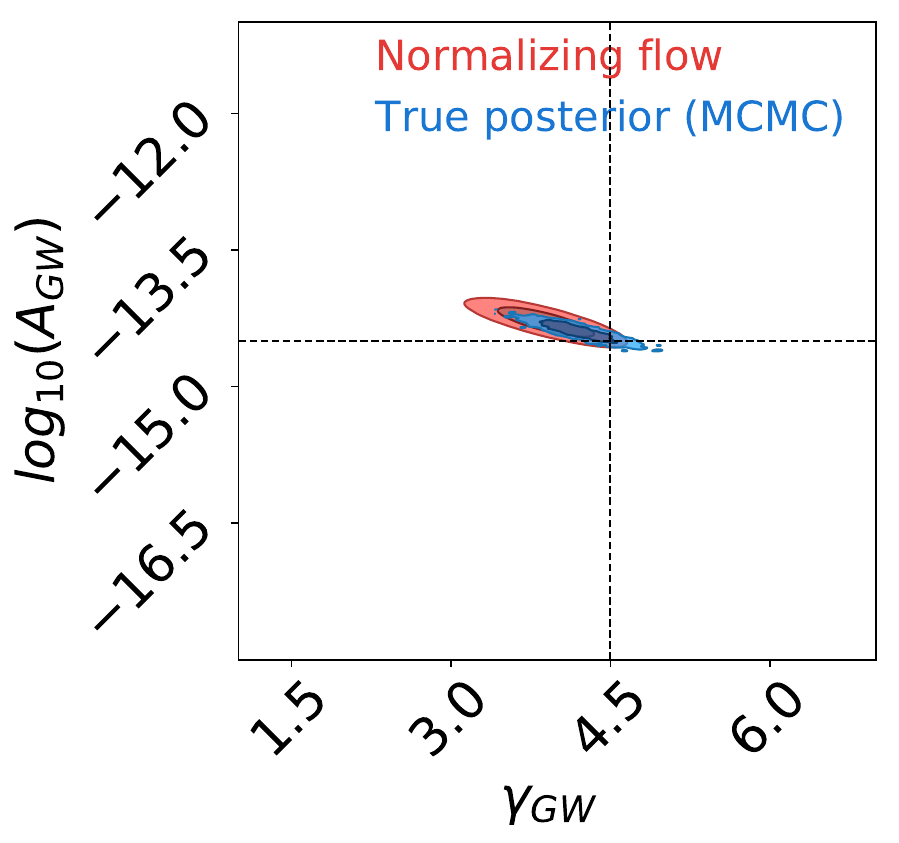}  
 \includegraphics[width=0.24\linewidth]{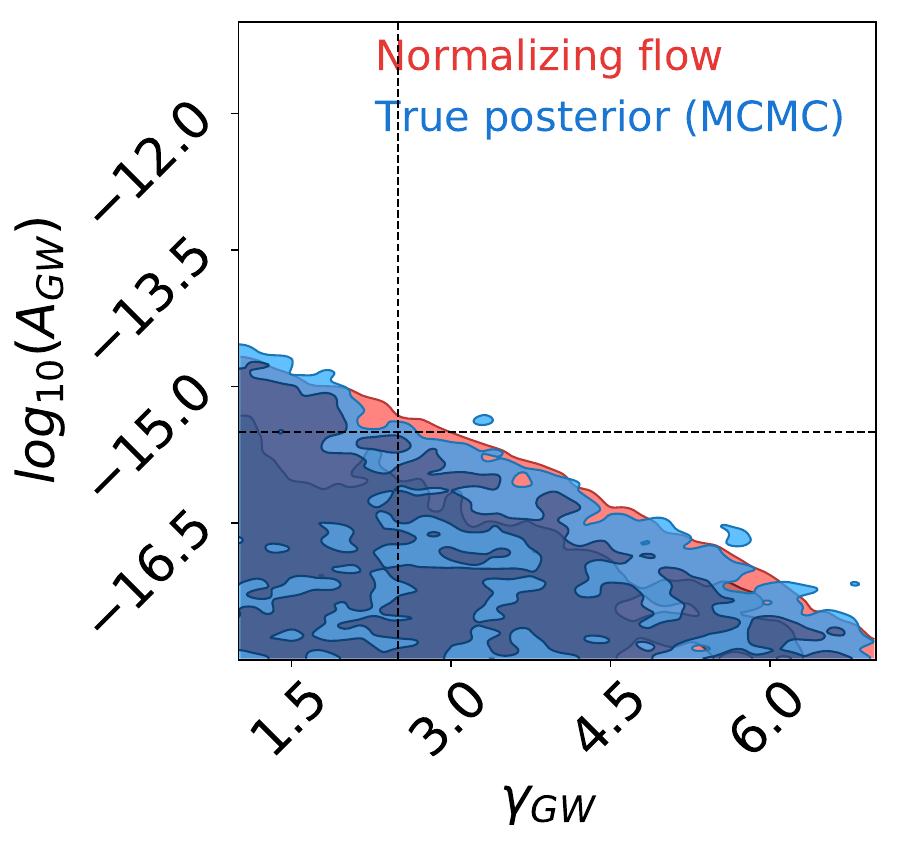} \includegraphics[width=0.24\linewidth]{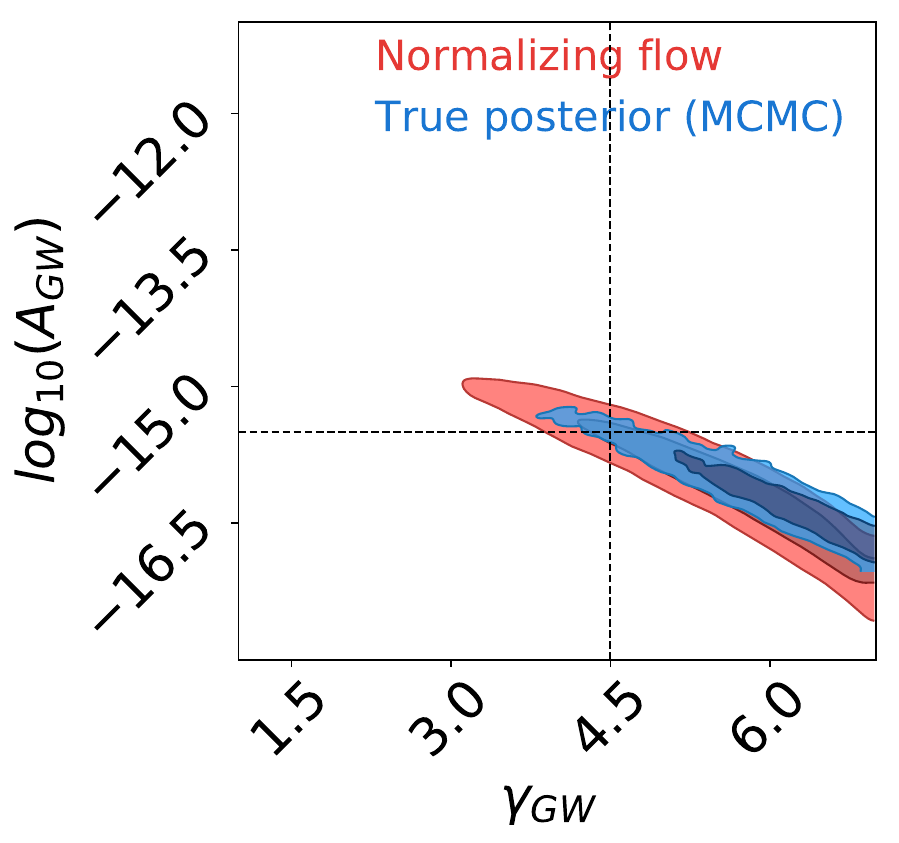}     
 \caption{A comparison of the flow-estimated posteriors and the MCMC-derived ground truth posteriors, for a grid of four different pairs of SGWB parameters: $(\gamma_{\rm GW},\log_{10}A_{\rm GW})=(2.5,-14.5),\,\, (4.5,-14.5),\,\, (2.5,-15.5),\,\, (4.5,-15.5)$ from left to right, respectively.} 
  \label{fig:posteriors_grid}
\end{figure*}

\textbf{Results.} We first show in \cref{fig:posteriors_grid} the posteriors recovered by our normalizing flow for a $2 \times 2$ grid of $(\log_{10}A_{\rm GW},\gamma_{\rm GW})$ values.  For each parameter choice, a single instance of pulsar residuals is generated and fed as conditional labels to the trained flow, and 100k samples in parameter space are generated from the flow. (In this example and all the subsequent ones, we fix the injected red noise values to their nominal best fit values shown in \cref{tab:pulsarinfo}.) We use \texttt{chain\_consumer} \cite{Hinton2016} to plot the posteriors from the samples. In \cref{fig:posteriors_grid}, we also show posteriors obtained from the exact PTA likelihood, using pulsar dataset simulations passed through the \texttt{ENTERPRISE} \cite{ellis_justin_a_2020_4059815} PTA data analysis pipeline, and sampled using MCMC. We can see that the flow-generated posteriors are already quite accurate, matching the true MCMC posteriors reasonably well across the parameter space. In particular, the flow posteriors more or less cover the MCMC ones, and correctly indicate when the SGWB parameters can be recovered vs.\ when the amplitude or the slope are too small and the posterior corresponds to only an upper limit.\footnote{Indeed, the flow correctly reports that the posterior in these cases carries no information below an approximately diagonal line in $\log_{10}(A_{\rm GW})$ vs.\ $\gamma_{\rm GW}$. This is expected from the nature of the PTA, whereby the bulk of the constraint on the SGWB comes from the lowest frequencies, and there one can trade off amplitude for slope as indicated in the third panel of \cref{fig:posteriors_grid}.}

\begin{figure*}[t]
  \centering
  \includegraphics[width=0.24\linewidth]{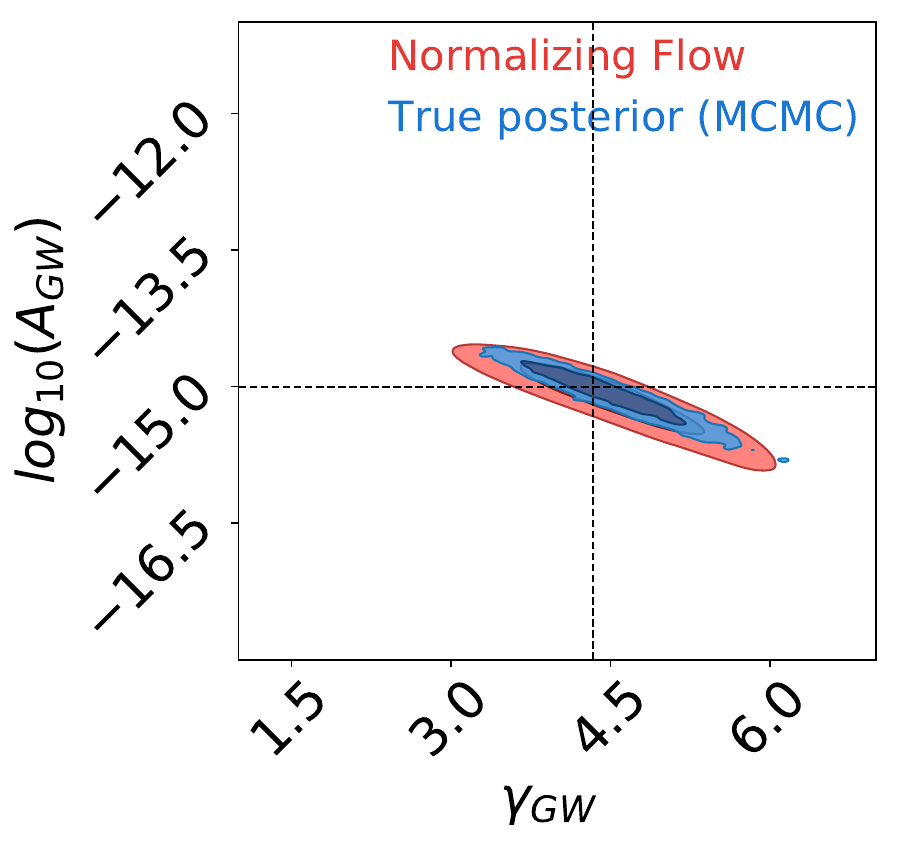}      \includegraphics[width=0.24\linewidth]{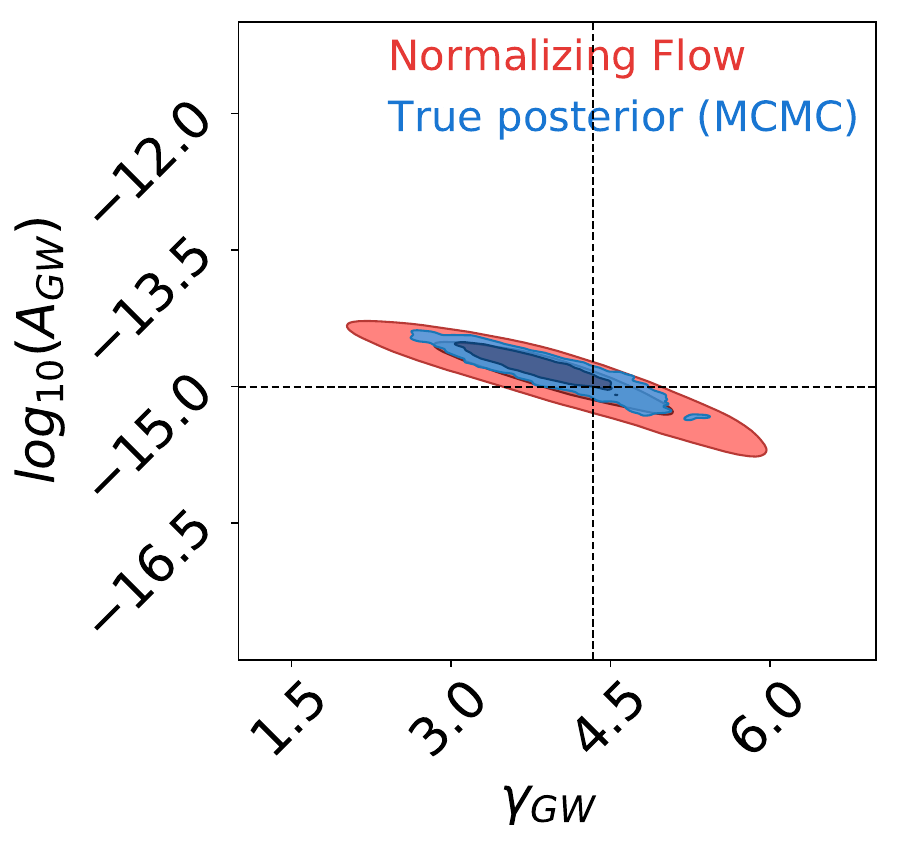} 
 \includegraphics[width=0.24\linewidth]{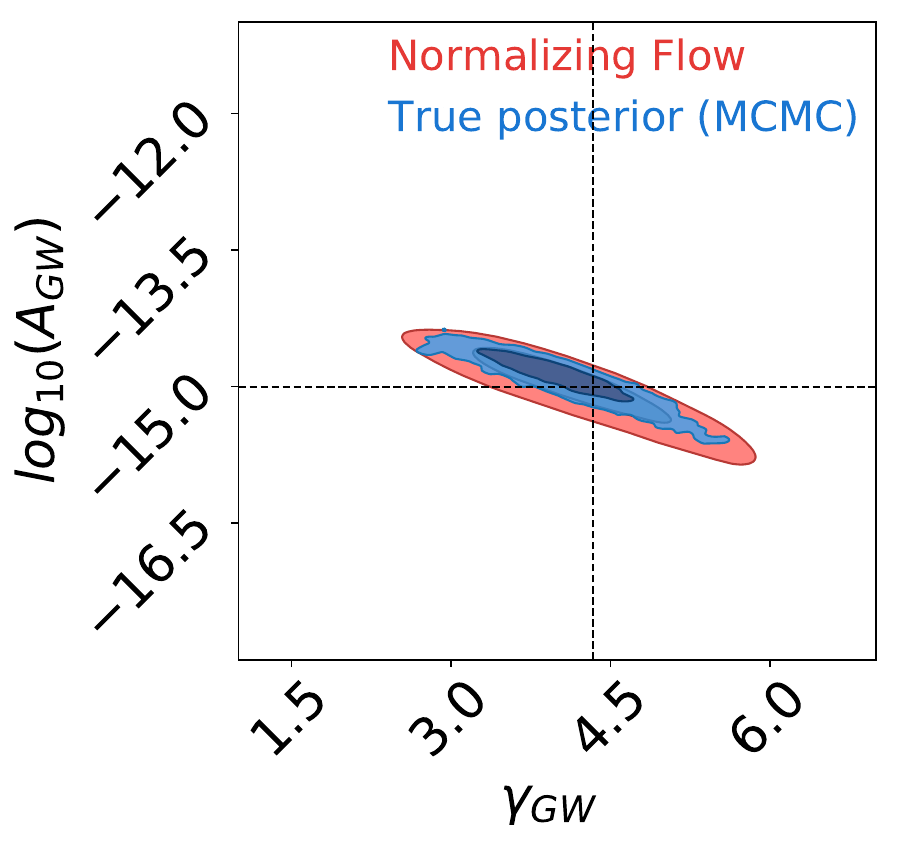}    \includegraphics[width=0.24\linewidth] {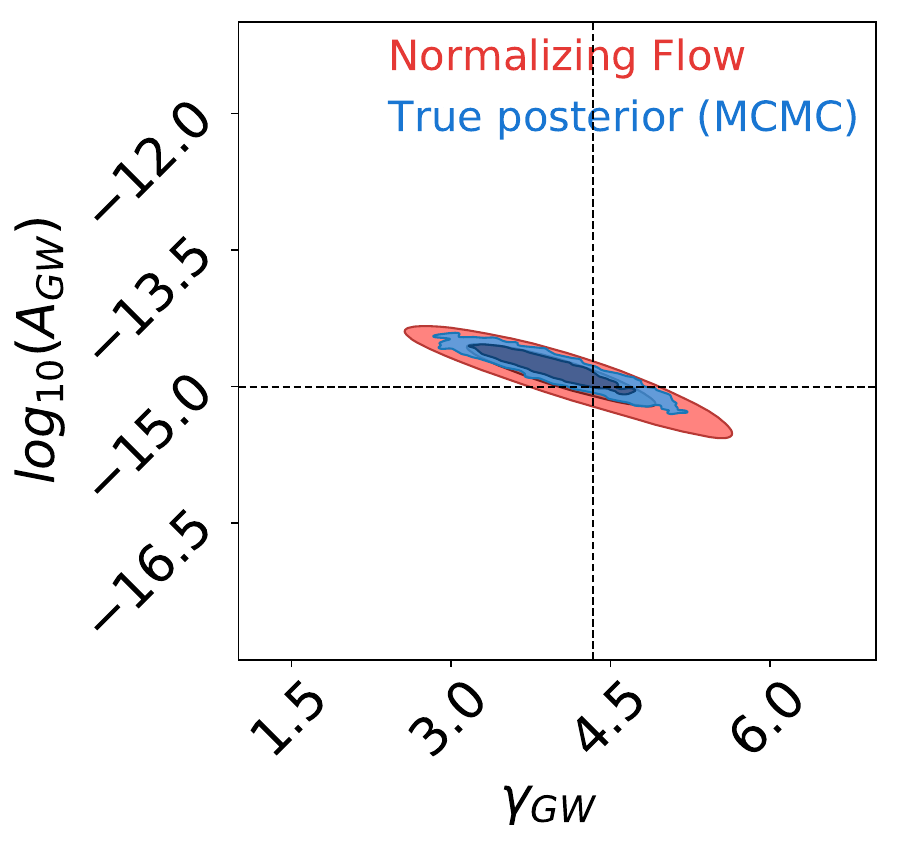}     
 \caption{A comparison of the flow-estimated posteriors and the MCMC-derived ground truth posteriors, for four time series sampled from a single choice of SGWB parameters, $(\gamma_{\rm GW},\log_{10}A_{\rm GW})=(13/3,-15)$.
 }
 \label{fig:posteriors_nominal}
\end{figure*}

Next, we drill down to the case of nominal SGWB values of $\gamma_{\rm GW}=13/3$ and $A_{\rm GW}=10^{-15}$. We can repeatedly generate timing residuals with these SGWB parameters and see the variation in posteriors that result. Four instances are shown in Fig.~\ref{fig:posteriors_nominal}. We again observe qualitatively good coverage of the true (MCMC) posterior.  We quantify the agreement with the true posteriors using  the Hellinger distance~\cite{Hellinger1909}. This is a distance measure between probability distributions which becomes tractable when the distributions are Gaussian -- an approximation that empirically describes our two-dimensional posteriors quite well. We find that the mean and standard deviation in Hellinger distances between flow and true posteriors calculated across 10 instances is $0.33\pm 0.04$.

Finally, the flow-generated samples can be made more precise by reweighting them with the true likelihoods. This reweighting technique has been explored previously in the PTA literature by \citet{Hourihane:2022ner}, who studied the efficacy of reweighting an approximate posterior obtained by ignoring cross correlations between pulsars. (See also \cite{Dax:2022pxd} who explored a very similar reweighting technique starting from flow-based posteriors, in the context of gravitational wave interferometry.)
Given a parameter point $\theta_a \sim p_\mathrm{flow}(\theta|r)$ sampled from the flow, we can calculate the true likelihood of these parameter points $p_\mathrm{true}(r|\theta_a)$. By Bayes' theorem, this likelihood is related to the true posterior by the prior $p(\theta_a)$ (which is taken to be uniform here), and an overall $\theta_a$-independent normalization (the Bayesian evidence):
\begin{equation}
p_\mathrm{true}(r|\theta_a)=\frac{p_\mathrm{true}(\theta_a|r)p(r)}{p(\theta_a)} \,.
\end{equation}
The Bayesian evidence $p(r)$ is independent of the model (flow or otherwise) that we sampled from. Up to this overall normalization, we can determine the weights required to turn the flow samples into samples following the true posterior:
\begin{equation}
  w_a = \frac{p_\mathrm{true}(r|\theta_a)p(\theta_a)}{ p_\mathrm{flow}(\theta_a|r)}.
\end{equation}
Furthermore, the average of the weights provides an estimate for the Bayesian evidence:
\beq
p(r) = \frac{1}{N} \sum_{a=1}^N w_a.
\eeq
We can use this estimate of $p(r)$ as a metric to compare two different models; if the importance sampling with a given $N$ does not accurately approximate the integral over the true posterior, the estimated evidence will be biased to lower values \cite{Dax:2022pxd}.

Another common measure of the quality of importance sampling (used, e.g., \ in \cite{Hourihane:2022ner,Dax:2022pxd}) is the \emph{weighting efficiency} (which is closely related to the \emph{effective sample size} \cite{KongESS}) 
\begin{equation}
  \varepsilon = \frac{1}{N} \frac{\left(\sum_{a=1}^N w_a\right)^2}{\sum_{a=1}^N w_a^2}\,.
\end{equation}
This is a statistical measure of the fraction of unweighted samples that would be required to match the variance of the weighted sample.

\begin{figure}[t]
  \centering
  \includegraphics[width=0.95\linewidth]{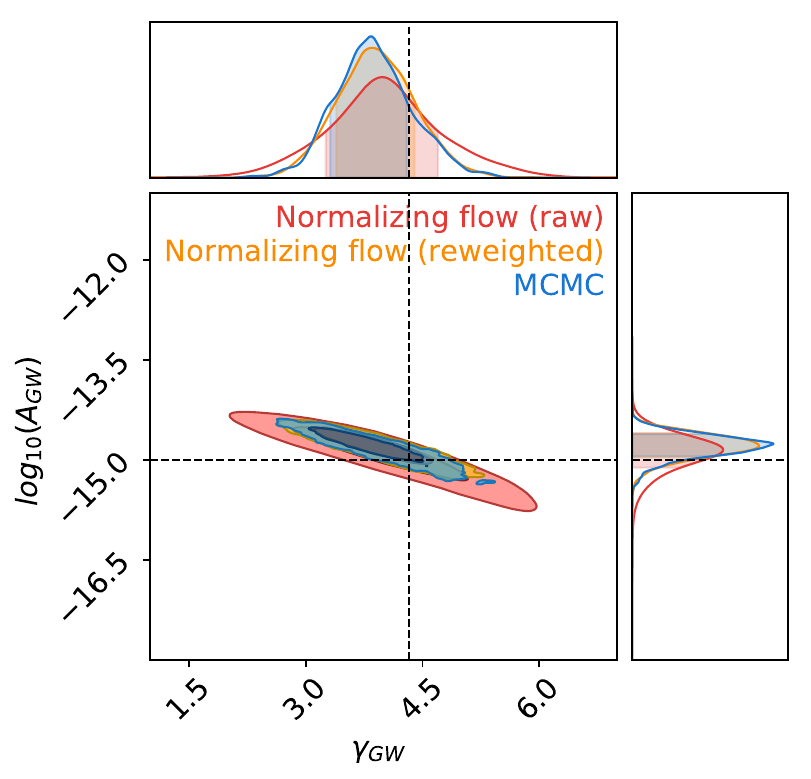}
  \caption{Posteriors from the raw normalizing flow (red), the reweighted normalizing flow (orange), and the MCMC-derived ground truth (blue), for a single realization of the PTA residuals with the same nominal SGWB parameters as in Fig.~\ref{fig:posteriors_nominal}.}
  \label{fig:posteriors_rw}
\end{figure}

The reweighted flow posteriors are shown in \cref{fig:posteriors_rw} compared with the MCMC posteriors for a single instance of timing residuals generated from the nominal SGWB parameters. We see the reweighted posteriors are significantly improved over the posteriors sampled from the uncorrected flow, basically in perfect agreement with the MCMC. 

With the reweighted posteriors, the Hellinger distances (again, calculated over 10 instances) improve to $0.22\pm0.15$. The effective sample sizes are estimated (using $N=10^6$ samples) to be $\log_{10}(\varepsilon)=-3.2\pm0.7$. While $\varepsilon\sim 0.1\%$ indicates there could definitely be room for further improvement, we note that it is still orders of magnitude better than using the prior as a proposal distribution (i.e., importance sampling with uniform distribution on parameter space). This gives $\log_{10}(\varepsilon)=-5.7\pm0.2$ for $N=10^6$, i.e., an effective sample size of essentially just one event, which basically means the importance sampling failed to provide any meaningful information about the true posterior. (Also, reweighting the uniform distribution gives a posterior that is in visual disagreement with the true posterior.)
Finally, the difference in estimates for the Bayesian evidence across 10 instances is 
\begin{equation}
  \log_{10} p(r)\Big|_\mathrm{NF} - \log_{10} p(r)\Big|_\mathrm{Uniform}
    = 9.7 \pm 0.7\,.
\end{equation}
We conclude the sum of the weights for the uniformly sampled distribution is nearly 10 orders of magnitude smaller than that of the flow sampled distribution.

Although reweighting the flow posteriors also requires evaluating the true likelihoods, it is an interesting alternative to the MCMC, for several reasons. First, these calls are fully uncorrelated and hence fully parallelizable, whereas the MCMC samples always suffer from some correlation and need to be evaluated (at least to some degree) sequentially. Second, the specific numbers and comparisons presented here are not set in stone --- the quality of the flow can likely be improved systematically with additional improvements to the neural network architecture. This will reduce the number of likelihood evaluations required for the reweighting, further improving the comparison with the MCMC (which is a stable, mature technique).

\textbf{Conclusions.} We have shown, using realistic simulations of PTA data, that normalizing flows offer enormous potential to vastly accelerate PTA data analysis and parameter inference with almost no loss in accuracy or sensitivity. Going forward, we expect these techniques enabled by modern machine learning to revolutionize the PTA field, complement the traditional MCMC-based techniques, and, with the influx of high-cadence data and new pulsars from forthcoming flagship radio facilities, ultimately replace the status quo pipelines. Machine-learning techniques, like the one we have studied here, will safeguard the future tractability and scalability of nanohertz-frequency GW analyses, ushering in a new era of discovery with PTA data.

There is much that can be improved in our pilot study. On the purely ML side, the specific architecture taken here (MAF-RQS normalizing flow with LSTM-based embedding) was not heavily optimized for performance, and it is likely that with a more dedicated hyperparameter scan, the performance of the flow-based posterior estimation could be greatly improved. It would also be fruitful to explore different architectures, e.g.\ embeddings based on transformers, or more expressive alternatives to ordinary normalizing flows such as diffusion models~\cite{sohldickstein2015deep, song2020generative_estimatingGradients, song2020improved_technieques_for_sorebased_geneerative, ho2020denoising, song2021scorebased_generativemodelling} or continuous normalizing flows~\cite{chen2019neural_CNF}. The results shown here should not be taken as the ultimate limit of what modern ML techniques can achieve, but just the starting point.

Pulsar timing data is highly heterogeneous in quality and regularity, as the limitations of legacy data are joined with the ever-improving sensitivity of modern data. Furthermore, there are many processes associated with the propagation of radio pulses in the ionized interstellar medium that leave their imprint on pulsar-timing data, and it is known that a one-size-fits-all approach to modeling these effects in pulsars is not appropriate. Machine-learning strategies must be able to accommodate the rich variety of noise processes with which pulsar-timing data must contend, and be able to do so on a per-pulsar basis. Additionally, GW signals in the PTA band are a combination of stochastic (e.g., the GW background) and deterministic (e.g., individually resolvable binary signals, or bursts), and machine-learning pipelines need to be able to model these with the same or better flexibility as current likelihood-centered approaches. Perhaps the most important improvement that must be made is the fact that the datasets are continually growing. Ideally a neural network would not need to be completely retrained to incorporate the extension of existing datasets, or indeed their expansion with additional pulsars.

Out of the improvements we have identified, there also lie opportunities. Simultaneous characterization of a GW background and a (perhaps variable) number of single resolvable GW signals remains challenging for current pipelines. Iterative refinement of pulsar noise models is also time consuming and somewhat ad hoc, in that it may depend on the assumed base model from which iteration is begun, and it currently does not take place at the level of the full array but rather independently in each pulsar. Deep learning, while not a panacea, could be well placed to tackle such complicated, high-dimensional decisions. If so, the discovery potential of PTAs will continue to grow, belying the long-timescale nature of the experiment to offer regular GW, pulsar, and interstellar-medium breakthroughs.

\begin{acknowledgments}

We thank Gabriel Freedman and members of the NANOGrav Detection Working Group for their useful comments on a preliminary form of this work. DS is supported by DOE grant DOE-SC0010008. MF was supported by the
DOE under grant DE-SC0010008 and the NSF under grant PHY1316222. SRT acknowledge supports from NSF AST-2007993, the NANOGrav NSF Physics Frontier Center \#2020265, and an NSF CAREER \#2146016. JD was supported in part by U.S. Department of Energy grant number DE-SC0023093. NS is supported in part by the NSF GRFP under Grant No. DGE-1842400. This work was conducted in part using the resources of the Advanced Computing Center for Research and Education (ACCRE) at Vanderbilt University, Nashville, TN.
\end{acknowledgments}

\appendix

\section{PTA dataset simulation methods}\label{app:homebrew}

\subsection{Time series generation}

Here we review the custom implementation we developed for simulating the PTA datasets used in training and benchmarking our conditional normalizing flows. While standard software frameworks, such as \texttt{libstempo} (built on \texttt{tempo2} \cite{2006MNRAS.369..655H,2006MNRAS.372.1549E}) and \texttt{PINT} \cite{2019ascl.soft02007L}, exist for PTA data simulation and analysis, we found them mismatched for our purposes. They are written primarily for the purpose of generating small amounts of data (on the order of existing datasets) with detailed implementations of perturbations on pulsar timing models from solar system and pulsar dynamics. In contrast, we require $O(\num{e6})$ datasets for training but are satisfied with a simplified treatment of effects which do not directly contribute to the stochastic structure of timing residuals in order to demonstrate the viability of our method. We therefore created a custom stripped-down model for the simulation of pulsar timing residuals from a combination of the SGWB and intrinsic pulsar noise, partially modelled on \texttt{libstempo}, fully vectorizing the code in \texttt{numpy} to create many similar datasets efficiently.

Both the SGWB and intrinsic pulsar noise are modeled as stationary time series, i.e., random Gaussian processes defined completely by their power spectra. 
A stationary spectrum is expected from the superposition of many individual monochromatic GWs that are too weak to be individually resolved on Earth. Individual pulsar noise is taken to be stationary, with the ansatz that it is characterized by a red power spectrum and a white noise term. Under these assumptions, the only difference between our signal and background is that the SGWB has a common-spectrum index and amplitude across the pulsar set and exhibits correlated noise between different pulsars with magnitude set by the Hellings--Downs function.

Pulsar white noise is modelled by adding an independent sample drawn from $\mathcal{N}(0, \sigma^2)$, where $\sigma = \text{100 nsec}$, to each residual. The remaining contributions to the residuals from pulsar noise and SGWB are characterized by a time series
\begin{equation}\label{eq:Ftint}
  r_I(t) = \int_{f_L}^\infty \dd{f} \pqty{a_I(f) \cos 2\pi ft + b_I(f) \sin 2\pi ft},
\end{equation}
where $I=1,\dots,N_p$ indexes the pulsar, and the Fourier components are completely defined by their power spectrum
\begin{align}
  \ev{a_I(f)a_J(f')} = \ev{b_I(f)b_J(f')} &= P_{IJ}(f) \delta(f-f') \,, \\
  \ev{a_I(f)b_J(f')} &= 0 \,.
\end{align}
The red spectrum for both the cosine and sine coefficients is parameterized by
\begin{equation}\label{eq:redspec}
  P_{IJ}(f) = \frac{A_{\rm GW}^2}{12\pi^2} \frac{1}{f_0^3} \pqty{\frac{f_0}{f}}^{\gamma_{\rm GW}}\chi_{IJ}+
  \frac{A_{r,I}^2}{12\pi^2} \frac{1}{f_0^3} \pqty{\frac{f_0}{f}}^{\gamma_{r,I}}\delta_{IJ},
\end{equation}
where $\chi_{IJ}$ are the Hellings-Down correlation coefficients, and by convention $f_0 = \SI{1}{\per\year}$. The low frequency cutoff $f_L$ on the integral is necessary for the residuals to not diverge from the low frequency contribution. However, this cutoff is not physical and we need to ensure that its value does not produce a detectable effect on our simulated data.

We approximate this integral by discretizing it,
\begin{equation}
  r_I(t) \approx \sum_{k=0}^{N_f-1} \Delta f
                 \pqty{a_{Ik} \cos 2\pi f_k t + b_{Ik} \sin 2\pi f_k t} \,,
\end{equation}
with $f_k = f_L + k \Delta f$. Now in order to generate a time series, we simply need to sample $2N_fN_p$ variables $a_{Ik}, b_{Ik}$ drawn from normal distributions normalized such that \cref{eq:redspec} is satisfied.

We also need to select the constraints $f_L, \Delta f$, and $N_f$ so that the 
time series is well-approximated. We find that
\begin{equation}
f_L=\Delta f={0.01\over T_{\rm obs}}, \qquad N_f=1000\,,
\end{equation}
where $T_{\rm obs}\approx 12.5$~yr is the full observation length of the PTA dataset,
was sufficient to guarantee the accuracy of the approximation. The small spacing  $\Delta f$ guaranteed the integral (\ref{eq:Ftint}) was well approximated --- both the steeply falling power spectrum in (\ref{eq:redspec}) and the trigonometric coefficients would be well sampled by the discretization. The large value of $N_f$ guaranteed that the high-frequency tail would be sufficiently suppressed by the red power spectrum (\ref{eq:redspec}) and that truncating it would not affect the time series. Finally, as we will review in the following subsection, choosing $f_L T_{\rm obs}\ll 1$ was important to guarantee that the time-domain covariance matrix (to calculate the exact likelihoods) is not regulator dependent.

\subsection{Quadratic timing fit and $\mathsf{G}$-matrix projection}

In the previous subsection we introduced the signal present in the stochastic timing residuals, constructed by taking the time of arrivals (TOAs) and subtracting away a detailed, deterministic timing model for each pulsar. The parameters making up the timing model are typically a prior unknown but are estimated from fits carried out by previous analyses. Since a signal is typically correlated with the best-fit parameters, this fitting needs to be redone simultaneously with a gravitational wave search. 

We work with a minimal timing model for each pulsar consisting of a quadratic function of the TOAs $t_i$ such that the ``design matrix'', $M$, has elements, $M_{i1} = 1$, $M_{i2} = t_i$, and $M_{i3} = t_i^2$. The residuals are then given by, 
\begin{equation}
r_{\rm obs}(t_i) = r(t_i) + M_{in} \xi _{n} \,, 
\end{equation}
where $\xi_n$ is the difference of the $n$th timing model parameters from their best-fit values. \footnote{In this discussion, we focus on the case of a single pulsar for simplicity; the generalization to multiple pulsars is straightforward.}

We write the Gaussian likelihood as,
\begin{equation}
    {\cal L} = \frac{1}{\sqrt{|2\pi \mathsf{C}|}} \exp \left(-\frac{1}{2} ({\vec r}_{\rm obs}-M {\vec \xi})^T\mathsf{C}^{-1} ({\vec r}_{\rm obs}-M\vec{\xi})\right)\,.
\end{equation}
$\mathsf{C}$ is the time-domain covariance matrix of timing observations computed under the assumption of a given red noise power spectrum; and $\vec r_{\rm obs}=(r_{\rm obs}(t_1),r_{\rm obs}(t_2),\dots r_{\rm obs}(t_{N_{\rm TOA}}))$ is the vector of pre-fit timing residuals.

Following~\cite{vanHaasteren:2012hj}, we then marginalize the likelihood over $\xi$, assuming a uniform, infinite, prior. We perform a singular value decomposition of $\mathsf{M} = \mathsf{U} \mathsf{\Sigma} \mathsf{V}^\dagger$, where $\mathsf{U}$ and $\mathsf{V}$ are unitary matrices. Being unitary, its columns form an orthogonal basis, and separating them into $\mathsf{U} = \pmqty{\mathsf{F} & \mathsf{G}}$, where $\mathsf{F}$ is $N_\mathrm{TOA} \times 3$ and $\mathsf{G} = N_\mathrm{TOA} \times (N_\mathrm{TOA}-3)$, $\mathsf{G}$ forms a projection matrix of timing residual components that are invariant to the shift by the timing model. Marginalizing over $\vec \xi$ leads to a likelihood consistent of projected residuals, 
\begin{equation}
\vec r_\mathrm{proj} = \mathsf{G}^T \vec r_{\rm obs}\,.
\end{equation}
These are the values the normalizing flows are conditioned upon.

We are also finally in a position to explain the lack of sensitivity to the regulator $f_L$ of the Fourier transform. Due to the projection above, the effective covariance matrix of projected residuals is $\mathsf{G}^T \mathsf{C} \mathsf{G}$. By appropriately Fourier transforming \cref{eq:redspec}, one can show that in the limit $f_L\tau\ll 1$, the time-domain covariance function $C(\tau)$ breaks up into universal, regulator-independent terms starting at $\tau^{\gamma-1}$, and non-universal regulator-dependent terms that are a power series in $(f_L\tau)^2$. The $\mathsf{G}$-matrix projection can be shown to remove the dependence on the regulator up to quartic order in $\tau=|t_i-t_j|$. Therefore, as long as $f_L|t_i-t_j|\ll 1$ such that the power series expansion is valid, the regulator dependence of the effective covariance matrix will be subleading as long as $\gamma < 7$ (which was the upper limit of the prior assumed in this work). And by choosing $f_L T_{\rm obs}\ll 1$, we ensure that $f_L|t_i-t_j|\ll 1$ for every pair of residuals in the PTA dataset.

\bibliographystyle{apsrev4-1}
\bibliography{bib}% Produces the bibliography via BibTeX.

\end{document}